\documentclass[aps,amsmath,amssymb,amsfonts,twocolumn,pra]{revtex4}
\usepackage{graphicx}
\usepackage{bm}
\usepackage{color}

\newcommand{\mbf}{\mathbf}
\newcommand{\mrm}{\mathrm}

\begin{document}
\author{Zbigniew Idziaszek$^1$, Tommaso Calarco$^{2,3}$, Paul S. Julienne$^4$, and Andrea Simoni$^5$}
\affiliation{
$^1$Institute of Theoretical Physics, University of Warsaw,
00-681 Warsaw, Poland\\
$^2$Institute of Quantum Information
Processing, University of Ulm, D-89069 Ulm, Germany\\
$^3$ European Centre for Theoretical Studies in Nuclear Physics and
Related Areas, I-38050 Villazzano (TN), Italy\\
$^4$Joint Quantum Institute, NIST and
the University of Maryland, Gaithersburg, Maryland 20899-8423, USA\\
$^5$Institut de Physique de Rennes,  UMR 6251 du CNRS and Universit\'e de
Rennes, 35042 Rennes Cedex, France}

\title{Quantum theory of ultracold atom-ion collisions}
\begin{abstract}

We study atom-ion scattering in the ultracold regime. To this aim, an
analytical model based on the multichannel quantum defect formalism
is developed and compared to close-coupled numerical calculations.
We investigate the occurrence of magnetic Feshbach resonances focusing
on the specific ${}^{40}$Ca$^{+}$+ Na system. The presence of several
resonances at experimentally accessible magnetic fields should allow the
atom-ion interaction to be precisely tuned. A fully quantum-mechanical
study of charge exchange processes shows that charge-exchange rates
should remain small even in the presence of resonance effects. Most of our results can be cast in a system-independent form and are important for the realization
of the charge-neutral ultracold systems.
\end{abstract}

\maketitle

Advances in trapping, cooling, manipulation and readout of single
atoms and ions have led over recent years to a range of fundamental
as well as applied investigations on the quantum properties of such
systems. Nowadays, an increasing number of experimental groups
worldwide are starting experiments with combined charged-neutral systems
in various configurations \cite{Vuletic}. While the theory of atom-ion collisions is
well established for high collision energies~\cite{Bransden1992,Delos1981},
a theoretical description in the ultracold domain is still largely
missing.

This letter presents the first study of magnetic Feshbach resonances
and the first fully quantum study of the radiative charge exchange process
for ultracold atom-ion systems that includes effects of Feshbach and
shape resonances. Here we consider only two-body collisions in free space, a necessary
prelude to further studies incorporating effects of ion micromotion or trap confinement.
We develop a reliable yet manageable effective model of
atom-ion collisions by applying multichannel quantum defect
theory (MQDT) \cite{Seaton,Greene,Mies} based on the long range
ion-induced-dipole potential that varies as $r^{-4}$
at large ion-atom distance $r$ \cite{SeatonClark,Avdeenkov}.  This powerful tool has proven effective as a
few-parameter approach for describing scattering and bound states in electron-ion
core \cite{Seaton}, electron-atom \cite{Watanabe} and neutral atom systems
\cite{Gao}. Although the literature on the subject is rich, here we discuss some details of MQDT illustrating how it works in the ultracold domain, so we can reveal the new and interesting ultracold ion-atom physics. We adapt MQDT to the atom-ion realm, utilizing the analytical
solutions for the $r^{-4}$ asymptotic potential \cite{Watanabe,Vogt} and
applying the frame transformation \cite{Burke,Gao} at short distances to
reduce the number of quantum defect parameters in the model.
We verify the model predictions by comparing to our own numerical close-coupled
calculations, taking ${}^{40}\mrm{Ca}^{+}\mrm{-}{}^{23}\mrm{Na}$ \cite{Makarov}
as a reference system.

We describe the $S$-state atom and $S$-state ion collisions with the
close-coupled radial Schr\"odinger equation
\begin{equation}
\label{RadialSchr}
\frac{\partial^2 \mbf{F}}{\partial r^2} + \frac{2 \mu}{\hbar^2}\left[E - \mbf{W}(r) \right] \mbf{F}(r) = 0.
\end{equation}
Here, $\mu=m_i m_a/(m_i+m_a)$ denotes the reduced mass, $\mbf{W}(r)$
is the  interaction matrix, and $\mbf{F}(r)$ is the matrix of radial
solutions. The wave function for $N$ scattering channels reads
$\Psi_i(\mbf{r}) = \sum_{j=1}^{N} A_j Y_j(\mbf{\hat{r}}) F_{ij}(r)/r$
where $Y_j(\hat{\mbf{r}})$ denotes the angular part of the solution
for the scattering channel $j$, and the constant vector $\mbf{A}$ is
determined by the boundary conditions at $r \rightarrow \infty$. The
asymptotic channel states can be characterized by the hyperfine
quantum numbers: $f_1$,$m_{f_1}$ and $f_2$,$m_{f_2}$ for ion and atom
respectively, and by the angular-momentum quantum numbers $l$ and $m_l$
of the relative motion of the atom and ion centers of mass. We denote the
asymptotic channel states by $|\Psi_\alpha\rangle$, where $\alpha = \{f_1
f_2 m_{f_1} m_{f_2} l m_l\}$. In the presence of a magnetic field $B$,
the field-dressed channel states $\left|\Psi_\alpha(B)\right\rangle$
are linear combinations of the bare $(B=0)$ channel states
$\left|\Psi_\alpha(B)\right\rangle = \sum_{\alpha^\prime} Z_{\alpha
\alpha^\prime}(B) \left|\Psi_{\alpha^\prime}\right\rangle$. In the
asymptotic channel basis the interaction matrix is diagonal at large distances
\begin{equation}
W_{ij}(r) \stackrel{r \rightarrow \infty}{\longrightarrow} \left[ E_i^{\infty} +\frac{\hbar^2 l(l+1)}{2 \mu r^2} - \frac{C_{4}}{r^4} \right]\delta_{ij} +  O(r^{-6}),
\label{Was}
\end{equation}
where $E_i^{\infty}$ are the threshold energies for the channel $i$
including the hyperfine energies and Zeeman shifts, and $C_{4} = \alpha
e^2/2$ with $\alpha$ denoting the static dipolar polarizability of the
atom and $e$ is the ion charge. At typical distances $R_0$ where the
short-range exchange interaction takes place, the interaction matrix
becomes diagonal in the $IS$ representation, characterized by total
electron spin $\mbf{S} = \mbf{s}_1 + \mbf{s}_2$, the total nuclear spin
$\mbf{I} = \mbf{i}_1 + \mbf{i}_2$, the total hyperfine angular momentum
$\mbf{F}=\mbf{f}_1+\mbf{f}_2=\mbf{I}+\mbf{S}$, and its projection $M_F$
on the axis of quantization, where $\mbf{s}_1$, $\mbf{s}_2$ are electron
spin of ion and atom respectively, and $\mbf{i}_1$, $\mbf{i}_2$ denote
their nuclear spins, respectively. We label those channels by $\beta =
\{I S F M_F l m_l\}$.

With the long-range atom-ion one can associate the length scale $R^\ast
\equiv \sqrt{ 2 C_4 \mu /\hbar^2}$ and energy scale $E^{\ast} =
\hbar^2/\left[2 \mu (R^\ast)^2\right]$. The length $R^{\ast}$
can be also related to the position of the last node of the zero-energy
$s$-wave radial wave function for the scattering length $a=0$: $L= 2/\pi
R^{\ast}$. The maxima due to the centrifugal barrier occur at $r_\mrm{max}
= \sqrt{2}/\sqrt{l(l+1)} R^{\ast}$ and have heights of $E_\mrm{max} =
\frac14 l^2 (l+1)^2 E^{\ast}$. Hence $E^\ast$ determines
the contribution of higher partial waves in the atom-ion scattering.

{\em MQDT.} The basic idea of MQDT \cite{Seaton,Greene,Mies} is the
separation between the short-range and the long-range properties of the
scattering wave function. The short-range wave function is insensitive
to the total energy $E$, and in the case of atom-ion collisions also
to the relative orbital angular momentum \cite{Gao2}. On the contrary,
at large distances the wave function vary rapidly with $E$, as well
as other scattering quantities defined at $r \rightarrow \infty$. By
introducing a set of short-range quantum-defect parameters, MQDT allows
one to determine all scattering and bound-state properties in a wide range
of collision energies.

In MQDT one starts by choosing a set of reference potentials $\{V_j(r)\}$,
that reproduce the asymptotic behavior of the interaction matrix at
large distances: $V_j(r) \stackrel{r \rightarrow \infty}{\longrightarrow}
W_{jj}(r)$, but otherwise can be arbitrary. With the reference potentials
$V_i(r)$ one can associate a pair of linearly independent solutions
$\hat{f}_i(r)$ and $\hat{g}_i(r)$
\begin{align}
\label{fghat}
\hat{f}_{i}(r) & = \alpha_i(r) \sin \beta_i (r),\\
\hat{g}_{i}(r) & = \alpha_i(r) \cos \beta_i (r).
\label{fghat1}
\end{align}
The amplitude $\alpha_i(r)$ fulfills the inhomogeneous Milne equation:
$[d^2/dr^2 + k_i(r)^2] \alpha(r) = \alpha^{-3}(r)$ \cite{Milne}, with
the local wavevector $k_i(r) = \sqrt{2 \mu (E - V_i(R)}/\hbar$,
while the phase $d\beta_i/dr = 1/\alpha_i^2$. We impose WKB-like
boundary conditions at small distances: $\alpha_i(r)
\cong 1/\sqrt{k_i(r)}$, $\alpha_i^\prime(r) \cong 0$, which makes
the functions $\hat{f}_i(r)$ and $\hat{g}_i(r)$ weakly dependent
on energy. The exact solution to Eq.\eqref{RadialSchr} at large
distances (where $V_j(r) \cong W_{jj}(r)$) can be expressed in terms of
$\hat{f}_i(r)$ and $\hat{g}_i(r)$: $\mbf{F}(r) \stackrel{r \rightarrow
\infty}{\longrightarrow} \left[\hat{\mbf{f}}(r) + \hat{\mbf{g}}(r)
\mbf{Y}(E)\right] \hat{\mbf{A}}$. Here, $\hat{\mbf{f}}(r)
\equiv \mrm{diag}[\hat{f}_{i}(r)]$, $\hat{\mbf{g}}(r) \equiv
\mrm{diag}[\hat{g}_{i}(r)]$, and $\mbf{Y}(E)$ is the energy insensitive
($\mbf{Y}(E) \cong \mbf{Y}$) quantum-defect matrix that plays a central
role in the quantum-defect analysis.

{\em Analytical solutions.} The Schr\"odinger equation with the polarization potential $C_4/r^4$ can be solved analytically, in terms of Mathieu functions of imaginary argument \cite{Vogt,Spector}. For our choice of MQDT reference potentials: $V_i(r) = C_4/r^4 + l(l+1)/r^2 + E_i^\infty$, the solutions: $\hat{f},\hat{g}$ are singular at $r \rightarrow 0$ \cite{AtomIon}
\begin{align}
\label{fhatS}
\hat{f}(r) & \cong r \sin(-R^{\ast}/r + \varphi), \quad (r \rightarrow 0) \\
\label{ghatS}
\hat{g}(r) & \cong r \cos(-R^{\ast}/r + \varphi), \quad (r \rightarrow 0)
\end{align}
Here, $\varphi$ denotes some short-range phase, that can be related
to the $s$-wave scattering length $a=-R^{\ast} \cot \varphi$ of the
$C_4/r^4$ potential \cite{AtomIon}. In view of the arbitrariness of
the reference potential, we set $\varphi =0$, which does not affect the
physical scattering matrices, that are not sensitive to the short-range
behavior of the reference potentials.

{\em Frame transformation.}  The frame transformation (FT) is a unitary
transformation between channels $\alpha$ and $\beta$: $U_{\alpha\beta} =
(f_1 f_2 m_{f_{1}} m_{f_{2}}| I S F M_F)$, that can be expressed in terms of the usual Clebsch-Gordan and Wigner $9j$ symbols (see e.g. \cite{Gao}).
The application of FT requires $R_0 \ll  R^{\ast}$,
which is typically very well fulfilled in the atom-ion
collisions. At distances $r \gtrsim R_0$ the exchange interaction is
no longer present, and for $r \ll R^\ast$ we can safely neglect both
the centrifugal potential and hyperfine splittings, and also ignore the
higher order dispersion terms: $C_6/r^6,C_8/r^8$ that lead only to small
corrections (see below). Then for $R_0 \lesssim r \ll R^\ast$,  $W_{ij}(r)
\cong \delta_{ij} C_4/r^4$, and the wave function in each of the
channel is simply given by the linear combination of the two short-range
solutions \eqref{fhatS} and \eqref{ghatS}, both in the asymptotic and
($IS$) representations.

In view of the assumed form of the reference
potentials, and the choice $\varphi=0$, the quantum-defect matrix
in $IS$ representation takes the form $Y^{(IS)}_{\beta \beta^\prime}
= \delta_{\beta \beta^\prime}[a_{S(\beta)}]^{-1}$, where $S(\beta)$
is the total electron spin in the channel $\beta$, that takes values
$S(\beta)=0,1$ corresponding to the singlet $a_s$ and triplet $a_t$
scattering lengths, respectively. Then, $\mbf{Y}$ in the basis of
the asymptotic channel states reads $\mbf{Y}= \mbf{Z}(B) \mbf{U}
\mbf{Y}^{(IS)} \mbf{U}^\dagger \mbf{Z}^\dagger(B)$ where $\mbf{Z}(B)$ is the transformation from bare to the magnetic-field-dressed states. Finally, the $\mbf{S}$ and $\mbf{K}$ matrices describing all the scattering properties including the cross-sections, are calculated directly from $\mbf{Y}$ with the help of the quantum-defect functions \cite{Greene,Mies}, that are determined by the asymptotic behavior of the short-range normalized solutions  $\hat{f},\hat{g}$.

\begin{figure}
\includegraphics[width=8.6cm,clip]{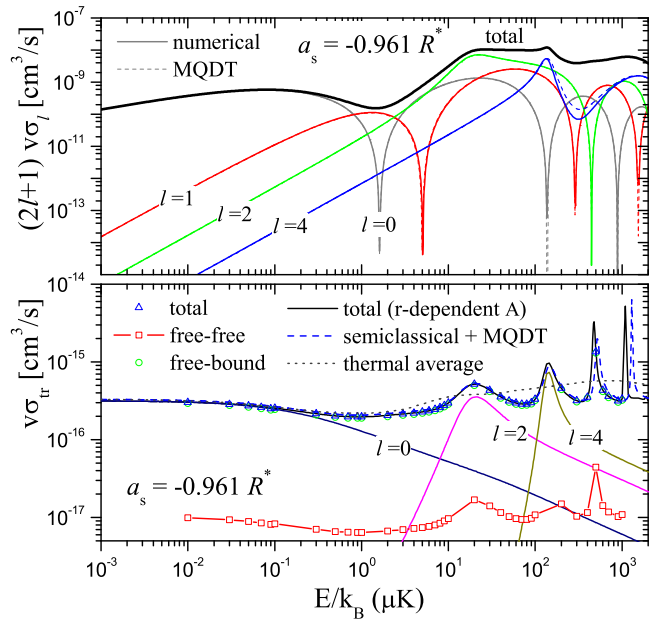}
\caption{
\label{Fig:Rates}
(Color online) Elastic collision rate in the singlet channel $A^{1} \Sigma^{+}$ (upper
panel) and rate of radiative $A^{1} \Sigma^{+} \to X^{1} \Sigma^{+}$
charge transfer (lower panel) versus collision energy. Different
theoretical approaches are compared (see text). Both the elastic and
the total inelastic rate are decomposed on selected partial waves.}
\end{figure}

{\em Results.} We now apply our model to a system of $^{40}\mrm{Ca}^{+}$
and $^{23}\mrm{Na}$ for which the long-range parameter $C_4 =
81.35$ a.u. and the short-range {\it ab-initio} potential curves are
approximately known \cite{Makarov}. The close-coupled Schr\"{o}dinger
equation \eqref{RadialSchr} is solved both within our MQDT approach and
exactly using standard numerical methods.

Unfortunately, potentials calculated by {\it ab-initio} methods are
usually not sufficiently accurate to predict the value of the associated
scattering lengths.  Therefore, for our calculation we will either
assume typical magnitudes of $a_s$ and $a_t$ on the order of $R^\ast$
or we vary them. It is interesting to note that for a $r^{-4}$
potential the probabilities to find a positive or a negative scattering
length are equal in contrast to van der Waals $r^{-6}$ interaction where
their ratio is 3:4 \cite{Gribakin}.

The $^{40}\mrm{Ca}$ ion has vanishing nuclear spin ($i_1 = 0$), whereas
the $^{23}\mrm{Na}$ atom has $i_2 = 3/2$, resulting in a total hyperfine
angular momentum $f_2=1$ and $f_2=2$. Neglecting small anisotropic spin
interactions, rotational invariance implies that $l$, $m_l$, and $M_F$
are conserved quantities. In our calculations we will consider collisions
in the $M_F=1/2$ block, and use the FT to obtain $\mbf{Y}$ in the asymptotic
channel representation. At $B=0$ the off-diagonal elements of $\mbf{Y}$
are proportional to $1/a_c = 1/a_s - 1/a_t$, where $a_c$ is an effective
scattering length characterizing the strength of channel coupling.

We begin our analysis by considering a single $A^1\Sigma^+$ potential as entrance
channel. The upper panel of Fig.~\ref{Fig:Rates} shows the rate $v \sigma_l$ of elastic Ca$^+$ + Na collisions as a function a collisions energy for different partial waves, with $v$ the relative atomic velocity and $\sigma_l$ the partial wave cross section. The calculation is performed using a potential parameterized by a scattering length of $a_s = -0.961 R^{\ast} = -2000 a_0$. One can observe several dips of the collisional rates, corresponding to Ramsauer minima. We note that because of the relatively low centrifugal barriers, partial waves with $l>0$ give a significant contribution already in the $\mu$K regime. At small energies the rates for $l>0$ behave asymptotically as $\sigma_l v \sim E^{3/2}$, in agreement with the threshold behavior typical of $r^{-4}$ potentials \cite{Note1}.For $l=4$ and at higher energies some discrepancies between the numerical and MQDT approach can be observed, resulting from breakdown of the assumption
of angular-momentum insensitivity of the quantum-defect matrix in the
analytical model. On the basis of the MQDT model, one can argue that the elastic collisional rates in the triplet channels, would be of the similar order, with the particular structure
of maxima/Ramsauer minima determined by the value of $a_t$.

The lower panel of Fig.~\ref{Fig:Rates} shows the total rate $v \sigma_{\rm tr}$ of inelastic $A^{1} \Sigma^{+} \to X^{1} \Sigma^{+}$ collisions associated with
radiative charge transfer (see \cite{Makarov} for details). The numerical
results (first three entries in the legend) have been obtained in the
distorted-wave Born approximation by summing contributions from all
free-free and free-bound transitions \cite{Julienne1}. Note that molecular ions are formed with very high probability ($\sim 96\%$) in the electron exchange process.
The figure also shows the total charge-exchange rate obtained by a
computationally much simpler single channel calculation with a position-dependent
Einstein coefficient $A(r)$ (see e.g. \cite{Julienne}). In addition we show the result of our
analytical model where the charge-transfer probability has been described
in the semiclassical theory, while MQDT provides a proper rescaling of
the wave function between reaction and asymptotic zones. One can observe
that the three numerical and analytical approaches give virtually the
same result for the total (free-free plus free-bound) charge-exchange
rate at low energy.  It is only at higher energies, in particular near
the peaks associated to shape resonances, that the MQDT model becomes
less accurate. This can again result from the modification of the
quantum-defect matrix $Y$ for higher $l$.

In contrast to the purely semiclassical result of Ref. \cite{Makarov},
both elastic and inelastic quantum cross sections appear to be strongly
affected by the presence of shape resonances up to mK regime; see
Fig.~\ref{Fig:Rates}. We have verified that this is a general behavior
not related to the particular choice of the singlet scattering length
assumed for the calculation. The influence of shape resonances can
also be observed in the thermally averaged rate which only begins to
saturate to the asymptotic value in the mK regime.

\begin{figure}[tb]
\includegraphics[width=8.6cm,clip]{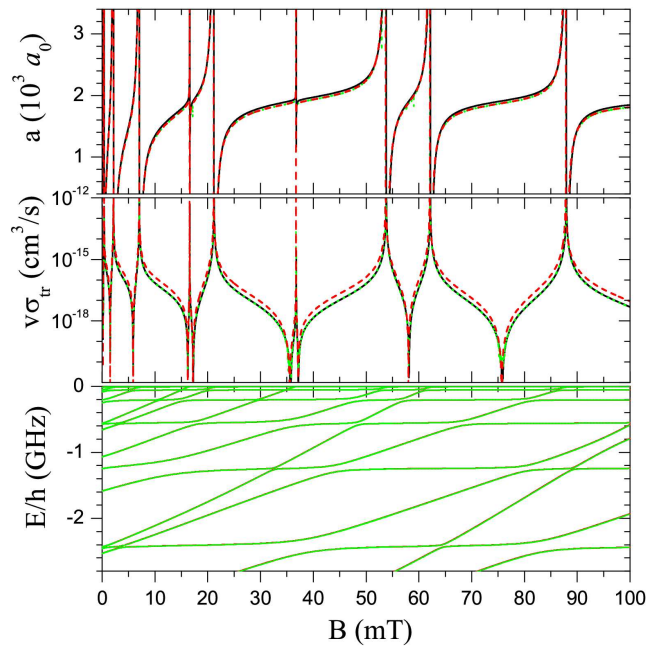}
\caption{
\label{Fig:Feshbach3}
(Color online) Field dependent $s$-wave scattering length (upper panel), charge transfer
rate for $s$-wave collisions at $E=1$nK (middle panel) and energies of $l=0$ bound states (lower panel) as a function of the magnetic field strength $B$, for
$a_s = - 0.9 R^{\ast}$ and $a_t  = -a_s$. Close-coupled numerical
calculations with (solid line) and without higher-order dispersion
terms (dotted line), and the MQDT results (dashed line) are compared.
Results from different approaches are undistinguishable on the scale of the
bottom panel.}
\end{figure}

Next, we take into account multichannel hyperfine effects and the presence of an external
magnetic field. Fig.~\ref{Fig:Feshbach3} shows the variation of the effective
scattering length $a$ versus magnetic field for sample values of the
singlet/triplet scattering lengths: $a_s = - 0.9 R^{\ast} = -1873 a_0$
and $a_t  = -a_s$. Several broad magnetic resonances can be observed in
the experimentally accessible range of magnetic fields. As for neutral
atoms, such resonances can in principle be used in practice to tune the atom-ion
effective interaction.

The MQDT results agree very well with the numerical results without
higher-order dispersion terms, whereas one can observe very small
discrepancies in the magnitude of $a$ in comparison to the calculations
performed with the full potential. Each resonance is associated with a
bound molecular level crossing the energetically lowest atomic threshold,
as shown in the lowest panel. The width of a resonance depends on particular molecular state
crossing the threshold. For small interchannel coupling $1/a_c$ bound states occur mainly in a single asymptotic channel $\alpha$, whereas for larger $1/a_c$ and weak magnetic fields, they are better characterized in terms of hyperfine angular momenta $\mbf{F},f_1,f_2$.

The middle panel compares the radiative charge-trans\-fer rate
calculated numerically with the closed-coupling method and a local
Einstein coefficient, and MQDT with the semiclassical description of the
short-range charge-transfer process. In the vicinity of resonances the
charge-transfer rate is strongly enhanced as expected, and the MQDT
model seems to slightly overestimate it.

The range of magnetic fields in which the first Feshbach resonances
should occur can be estimated from the energy location $E^\infty$ of the
associated bound states for $a=\infty$ and $B=0$. For a pure $r^{-4}$
potential such energies are $E^\infty/E^{\ast} = 105.8, 1180, 5208,...$
Letting $\Delta \mu$ be the difference of magnetic moments between the
Feshbach molecule and free atoms and assuming only linear Zeeman shift
for small values of $B$, we expect the first resonances to be found at
$B < B_\mrm{max}=E^\infty/\Delta \mu$, with $B_\mrm{max}[\textrm{mT}]= 0.086,
0.962, 4.26$ for Na-Ca$^{+}$ system.

\begin{figure}[tb]
\includegraphics[width=8.6cm,clip]{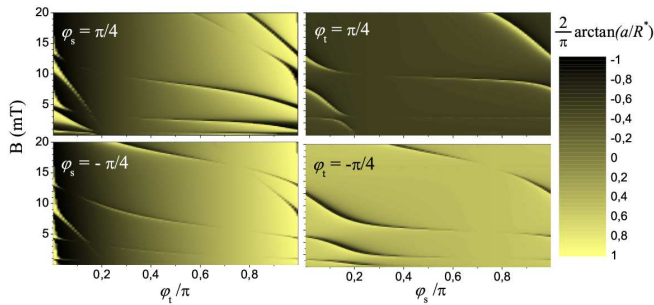}
\caption{
\label{Fig:Feshbach4} (Color online)
Contour plot of the dimensionless quan\-ti\-ty $(2/ \pi) \arctan(a/R^{\ast})$,
versus singlet and triplet short-range phases $\phi_s = - \arctan 1/a_s$
and $\phi_t = - \arctan 1/a_t$ and magnetic field $B$. In the
leftmost panels $\phi_s$ is fixed, in the rightmost panels $\phi_t$
is fixed.}
\end{figure}

In order to investigate the dependence of the position and
strength of Feshbach resonances on the interatomic potentials, we
show in Fig.~\ref{Fig:Feshbach4} the variation of the quantity $2
\arctan(a/R^{\ast})/\pi$ with the singlet and triplet short-range phases
$\phi_{s,t} = - \arctan 1/a_{s,t}$ and magnetic field $B$. Please note that typically several resonances should be observable below $B \sim 20$ mT. One can remark that the resonances are
usually relatively broad and only become very narrow when $a_s$ and $a_t$
are comparable, and the interchannel coupling $1/a_c$ is small. Apart
from the vicinity of resonances the field dependent scattering length is
mainly determined by the value of $a_t$, that again is in turn related
to the structure of the $\mbf{Y}$ matrix.

In conclusion, a relatively simple quan\-tum-de\-fect
mo\-del accurately describes atom-ion collisions  in the ultracold
domain, as verified by comparison
with numerical coupled-channel calculations for the reference system
${}^{40}$Ca$^{+}$ and ${}^{23}$Na. Our calculations predict that several
magnetic Feshbach resonances should be available at relatively
small values of magnetic fields to control the atom-ion interaction.
Radiative charge exchange rates remain relative
small even in the presence of Feshbach and shape resonances so that
elastic collisions can be controlled without introducing unwanted
losses.

We acknowledge support from the Office of Naval Research (PSJ), Polish Government Research Grant for 2007-2010 (ZI), European Commission through the Integrated Project FET/QIPC "SCALA" (TC), and National Science Foundation through a grant for the Institute for Theoretical Atomic, Molecular and Optical Physics at Harvard University and Smithsonian Astrophysical Observatory (ZITC,).

\end{document}